\documentclass[11pt]{article}
\usepackage{setspace}
\usepackage[left = 2cm, right=2cm, top=3cm, bottom=3cm]{geometry}
\usepackage[pdftex]{graphicx}
\usepackage{cite}
\usepackage{bm}
\usepackage{float}
\usepackage{amsmath,amssymb,latexsym}
\usepackage{color}
\usepackage[T1]{fontenc}
\usepackage{wasysym}
\usepackage{siunitx}
\usepackage{setspace}
\usepackage{relsize}
\usepackage{tikz}
\usepackage{lineno, blindtext}
\usepackage{authblk}
\usepackage[symbol]{footmisc}
\usepackage{lineno}

\begin{document}
	\title{\color{blue}\textbf{Side branched patterns, coalescence and stable interfaces during radial displacement of a viscoelastic fluid}}
	\author[1, $\dagger$]{Palak}
	\affil[1]{\textit{Soft Condensed Matter Group, Raman Research Institute, C. V. Raman Avenue, Sadashivanagar, Bangalore 560 080, INDIA}}
	\author[2, $\ddagger$]{ Rahul Sathyanath}
	\affil[2]{\textit{Department of Metallurgical and Materials Engineering, Indian Institute of Technology Madras, Chennai 600 036, India}}
	\author[2, $\mathsection$]{ Sreeram K. Kalpathy }
	\author[1,*]{Ranjini Bandyopadhyay}
	\vspace{0.5cm}
	\date{\today}
	\renewcommand{\figurename}{Fig.}
\maketitle

\footnotetext[2]{palak@rri.res.in}
\footnotetext[3]{mm15d023@smail.iitm.ac.in}
\footnotetext[4]{sreeram@iitm.ac.in}
\footnotetext[1]{Corresponding Author: ranjini@rri.res.in}
\renewcommand{\abstractname}{\vspace{-1cm}}

\begin{abstract}
	{\bf We explore the interfacial instability that results when a Newtonian fluid (a glycerol-water mixture, inner fluid) displaces a viscoelastic fluid (a dense cornstarch suspension, outer fluid) in a radial Hele-Shaw cell. As the ratio of viscosities of the inner and outer fluids is increased, side branched interfacial patterns are replaced by more stable interfaces that display proportionate growth and finger coalescence. We correlate the average finger spacing with the most dominant wavelength of interfacial instability, computed using a mathematical model that accounts for viscous fingering in miscible Hele-Shaw displacements. The model predictions on the role of viscosity ratio on finger spacing are in close agreement with the experimental observations. Our study lends insight into the significant contribution of the viscoelasticity of the outer fluid on the morphology and growth of interfacial patterns.}
\end{abstract}

	\definecolor{brass}{rgb}{0.71, 0.65, 0.26}
	\definecolor{junglegreen}{rgb}{0.16, 0.67, 0.53}
	\definecolor{indigo(web)}{rgb}{0.29, 0.0, 0.51}
	\definecolor{alizarin}{rgb}{0.82, 0.1, 0.26}
	\definecolor{red(ryb)}{rgb}{1.0, 0.15, 0.07}
	\definecolor{black}{rgb}{0.0, 0.0, 0.0}
	\definecolor{darkcandyapplered}{rgb}{0.64, 0.0, 0.0}
	\definecolor{darkred}{rgb}{0.55, 0.0, 0.0}
	\definecolor{darkolivegreen}{rgb}{0.33, 0.42, 0.18}
	\definecolor{internationalkleinblue}{rgb}{0.0, 0.18, 0.65}
	\definecolor{olivedrab(web)(olivedrab3)}{rgb}{0.42, 0.56, 0.14}
	\definecolor{saddlebrown}{rgb}{0.55, 0.27, 0.07}
	\definecolor{purple(munsell)}{rgb}{0.62, 0.0, 0.77}
	\definecolor{ufogreen}{rgb}{0.24, 0.82, 0.44}
	\newcommand{\bt}{\textcolor{brass}{$\blacktriangle$}}
	\newcommand{\brass}{\textcolor{brass}{\small$\ocircle$}}
	\newcommand{\BL}{\textcolor{black}{\small$\blacksquare$}}
	\newcommand{\blbullet}{\textcolor{black}{\large$\bullet$}}
	\newcommand{\darkred}{\textcolor{darkred}{$\blacksquare$}}
	\newcommand{\wine}{\textcolor{darkred}{\large$\bullet$}}
	\newcommand{\DR}{\raisebox{2pt}{\tikz{\draw[-,black!40!darkcandyapplered,solid,line width = 2.5pt](0,0) -- (5mm,0);}}}
	\newcommand{\hollow}{\textcolor{black}{$\square$}}
	\newcommand{\JG}{\textcolor{junglegreen}{\large$\varhexagon$}}
	\newcommand{\cyan}{\textcolor{junglegreen}{$\bigstar$}}
	\newcommand{\green}{\textcolor{darkolivegreen}{$\bigstar$}}
	\newcommand{\bluepentagon}{\textcolor{internationalkleinblue}{\large$\pentagon$}}
	\newcommand{\purplepentagon}{\textcolor{purple(munsell)}{\small$\triangle$}}
	\newcommand{\brown}{\textcolor{saddlebrown}{\large$\diamond$}}
	\newcommand{\olivedot}{\textcolor{olivedrab(web)(olivedrab3)}{$\square$}}
	\newcommand{\purpleline}{\raisebox{2pt}{\tikz{\draw[-,black!40!purple(munsell),solid,line width = 2.5pt](0,0) -- (5mm,0);}}}
	\newcommand{\indigo}{\textcolor{indigo(web)}{\large$\bullet$}}
	\newcommand{\blueline}{\raisebox{2pt}{\tikz{\draw[-,black!40!internationalkleinblue,solid,line width = 2.5pt](0,0) -- (5mm,0);}}}
	\newcommand{\oliveline}{\raisebox{2pt}{\tikz{\draw[-,black!40!ufogreen,solid,line width = 2.5pt](0,0) -- (5mm,0);}}}
	\newcommand{\candyred}{\textcolor{darkcandyapplered}{\large$\bullet$}}
	\newcommand{\blhex}{\textcolor{black}{\large$\varhexagon$}}
	\newcommand{\brsquare}{\textcolor{brass}{\small$\square$}}
	\newcommand{\brcircle}{\textcolor{brass}{\large$\circ$}}
	\newcommand{\brtria}{\textcolor{brass}{\small$\triangle$}}
	\newcommand{\brdown}{\textcolor{brass}{\small$\triangledown$}}
	\newcommand{\brdiamond}{\textcolor{brass}{\large$\diamond$}}
	\newcommand{\bltriangle}{\textcolor{internationalkleinblue}{$\triangleleft$}}
	\newcommand{\blrhd}{\textcolor{internationalkleinblue}{\large$\triangleright$}}
	\newcommand{\blcirc}{\textcolor{internationalkleinblue}{\large$\circ$}}
	\newcommand{\blpentagon}{\textcolor{internationalkleinblue}{$\pentagon$}}
	\newcommand{\bltri}{\textcolor{internationalkleinblue}{\small$\triangle$}}
	\newcommand{\blsq}{\textcolor{internationalkleinblue}{\small$\square$}}
	\newcommand{\rddiam}{\textcolor{darkcandyapplered}{\large$\diamond$}}
	\newcommand{\rdhexagon}{\textcolor{darkcandyapplered}{\small$\varhexagon$}}

\section{Introduction}
The displacement of a more viscous fluid by a less viscous fluid renders the fluid-fluid interface unstable and leads to intricate patterns called viscous fingers~\cite{Saffman,Homsy,Paterson_1981,Nature,Chen,Bensimon,Setu}. A fundamental understanding of these instability mechanisms is key to the design and enhancement of processes like oil recovery, sugar refining and carbon sequestration~\cite{Gorell,Hill,Orr,Cinar}. The radial Hele-Shaw cell is often used to track and obtain information about viscous fingering instabilities in a quasi two-dimensional flow~\cite{Hele-Shaw,Darcy}. A few studies have focused on the vital role that viscosity mismatch plays in the emergence of interfacial patterns~\cite{Lajeunesse,Irmgard}. In experiments by Lajeunesse $et$ $al$.~\cite{Lajeunesse}, the onset of instability at the interface between two miscible Newtonian fluids was studied. In spite of the low interfacial tension of the fluids, a complete suppression of instability was reported above a critical value of viscosity ratio, $\eta_{in}/\eta_{out}$, where $\eta_{in}$ and $\eta_{out}$ are, respectively, the viscosities of the inner and outer fluids. Bischofberger $et$ $al$.~\cite{Irmgard} studied the pattern evolution at the interface between a pair of miscible Newtonian fluids over a wide range of viscosity ratios. They reported a transition from a fractal growth regime (similar to that reported by Saffman and Taylor~\cite{Saffman}) to a stable regime (as reported by Lajeunesse $et$ $al$.~\cite{Lajeunesse}) as $\eta_{in}/\eta_{out}$ is increased.
\par
Owing to the complex flow behaviour of non-Newtonian fluids at finite shear rates, a detailed understanding of instabilities at the interface between a Newtonian and a non-Newtonian fluid remains elusive. Interfaces involving non-Newtonian fluids such as polymer solutions~\cite{Lindner_2000,Kawaguchi} and colloidal suspensions~\cite{Kagei,Lemaire} have been investigated in Hele-Shaw cells. A range of distinct pattern morphologies in the gas-driven displacement of cornstarch suspensions exhibiting discontinuous shear-thickening has been reported~\cite{Ozturk}. In a significant advance to earlier studies, we report here the emergence and growth of large scale patterns of distinct morphologies at a Newtonian/non-Newtonian liquid interface over a wide range of viscosity ratios. We observe and model the displacement of dense aqueous suspensions of cornstarch particles (outer fluid) by glycerol-water mixtures (inner fluid) in a radial Hele-Shaw cell. Transitions from side branched patterns to increasingly stable interfaces exhibiting proportionate growth and a new double coalescence phenomenon are reported. The concentration-dependent elasticity of the cornstarch suspension is seen to reduce the finger length, while resulting in the breakup of the interface into multiple small protrusions. The pattern morphologies at the initial stages are explained by employing a linear stability analysis of the interface between a Newtonian and a shear-thinning fluid~\cite{Azaiez}. The instability wavelength computed numerically shows close agreement with the average finger spacing estimated from experiments over three decades of viscosity ratios.

\section{Results}
	
\begin{figure}[!t]
\centering
\includegraphics[width=5.5in]{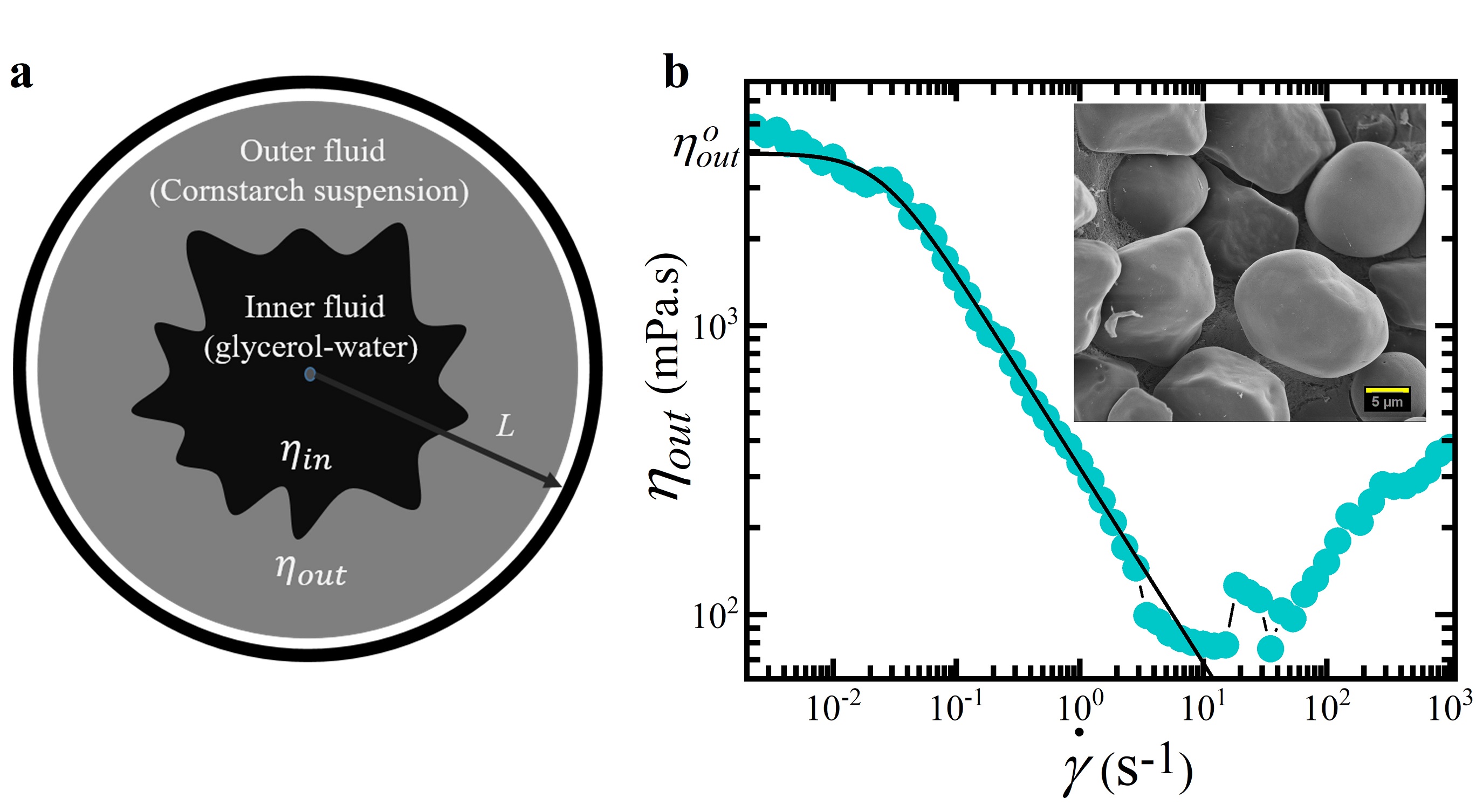}
\caption{$\vert${\bf Schematic of Hele-Shaw cell and flow curve of the cornstarch suspension.} {\bf a,} Schematic illustration of a radial Hele-Shaw cell (radius $L$ = 30 cm and gap between plates = 170 $\mu$m) in which the outer fluid with viscosity $\eta_{out}$ (shown by the dark gray peripheral region) is introduced through a circular hole (shown by the blue circle at the centre) on the top plate. The inner fluid with viscosity $\eta_{in}$ (shown by the black region) is next injected through the same hole. {\bf b,} Viscosity ($\eta_{out}$) $vs.$ shear rate ($\dot{\gamma}$) for 35 wt.\% aqueous cornstarch suspensions. The black solid line is the fit of the data to the simplified Carreau model, $\eta_{out} = {\eta_{out}^{o}} [(1+(\tau\dot{\gamma})^2)]^{(n-1)/2}$. Here, $\eta_{out}^{o}$ is the zero-shear viscosity of the cornstarch suspension, $\tau$ denotes its characteristic relaxation time and $n$ is an exponent. The extrapolated value of viscosity at a shear rate approaching zero is denoted by $\eta_{out}^{o}$ and has the value 3.952 Pa.s. The other fitting parameters are $\tau=41$ s and $n=0.32$. Inset shows a cryo-SEM image of cornstarch particles. The scale bar is 5 $\mu$m.} 
		\label{Viscosity-EP}
\end{figure} 

\begin{figure}[!t]
\centering
\includegraphics[width=5.5in]{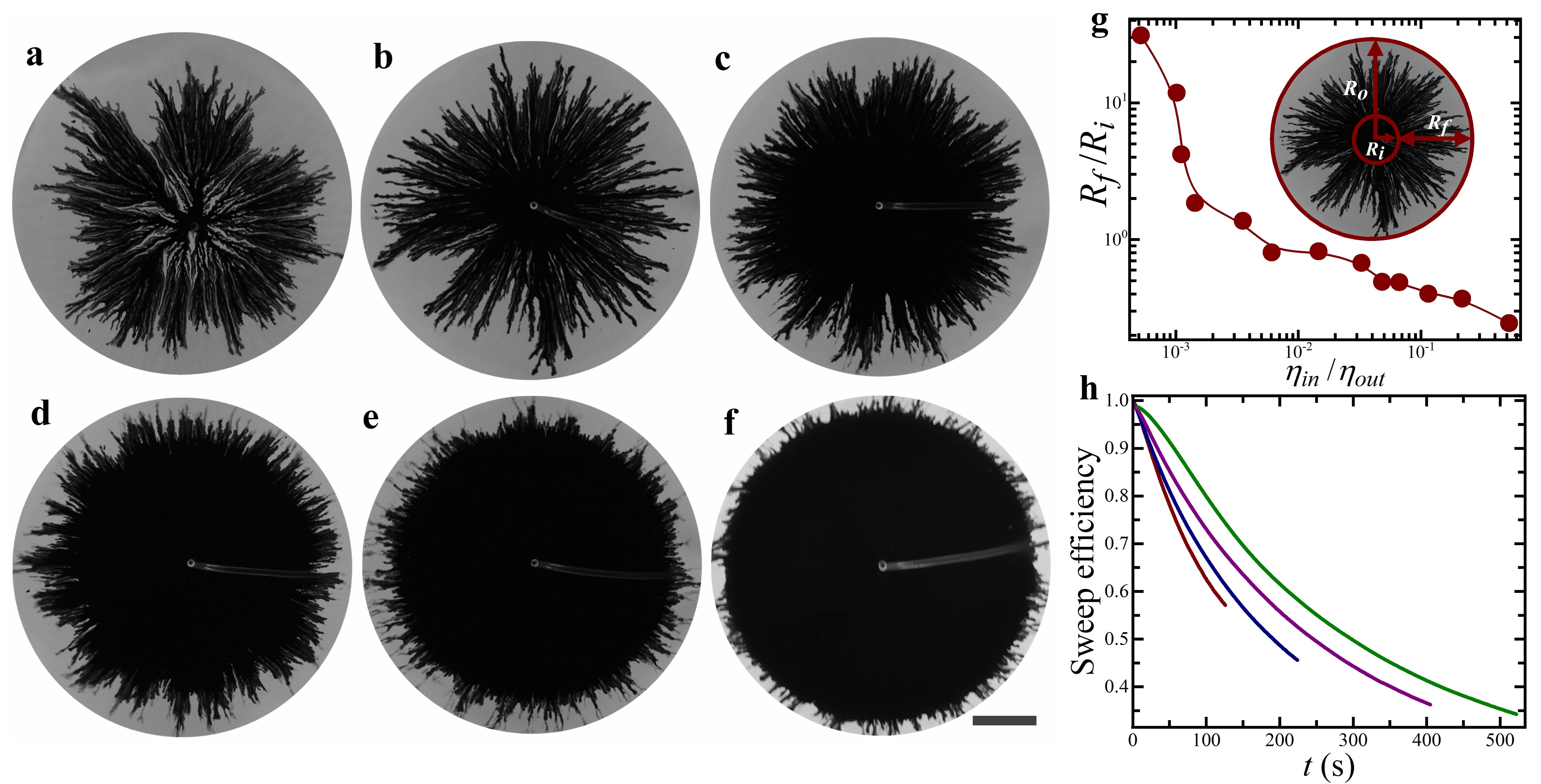}
\caption{$\vert${\bf Interfacial patterns with viscosity ratio as a control parameter.} {\bf a-f,} Patterns for increasing viscosity ratios: $\eta_{in}/\eta_{out}=\num{5.2e-4}, \num{1.1e-3}, \num{3.0e-3}, \num{3.0e-2}, \num{2.0e-1}, \num{5.0e-1}$ when a dyed glycerol-water mixture invades a 35 wt.\% cornstarch suspension in a radial Hele-Shaw cell. The scale bar is 5 cm. {\bf g,} The size ratio $R_f/R_i$ $vs.$ $\eta_{in}/\eta_{out}$ for $R_o$ = 13.5 cm. Inset shows the three characteristic length scales, $R_o$, $R_i$ and $R_f$. {\bf h,} Sweep efficiency $vs.$ time ($t$) for different viscosity ratios: $\eta_{in}/\eta_{out}=\num{5.2e-4}$ (\protect\DR), $\num{3.0e-3}$ (\protect \blueline), $\num{3.0e-2}$ (\protect \purpleline) and $\num{1.0e-1}$ (\protect \oliveline).}
\label{sizeratio-EP}
\end{figure}

Figure~\ref{Viscosity-EP}a shows a schematic of the Hele-Shaw cell (radius  $L$ = 30 cm, gap between plates $b$ = 170 $\mu$m), used for studying interfacial patterns. Figure~\ref{Viscosity-EP}b shows the non-monotonic viscosity profile of 35 wt.\% aqueous cornstarch suspensions with increasing shear rates. For shear rates upto $30$ $\mathrm{s^{-1}}$, shear-thinning behaviour is observed which fits a simplified Carreau model~\cite{Bird} as noted in the figure caption. The irregular-shaped cornstarch particles have a characteristic size of approximately 15 $\mu$m (inset of Fig.~\ref{Viscosity-EP}b). It has been proposed that shear-thinning cannot be attributed to particle rearrangement alone, but also to a constant hydrodynamic viscosity contribution arising from viscous stresses and an entropic contribution triggered by random particle collisions~\cite{Cheng,Crawford}. Beyond a critical shear rate, inter-particle frictional forces lead to the formation of a fragile soft solid which exhibits shear-thickening~\cite{Fall,Nagel,Peters}. In this work, shear-thinning cornstarch suspensions are displaced by glycerol-water mixtures in the radial Hele-Shaw geometry and the formation of interfacial patterns is recorded at different viscosity ratios of the two fluids. The viscosity ratio, $\eta_{in}/\eta_{out}$, is controlled by changing only $\eta_{in}$ by carefully varying the glycerol concentration in the glycerol-water mixture. The viscosity of the cornstarch suspension that corresponds to the shear rate ($\dot\gamma$ = 0.08 $s^{-1}$) imposed by the glycerol-water mixture at the time of injection is $\eta_{out}$. Representative interfacial patterns are shown in Figure~\ref{sizeratio-EP}. At low $\eta_{in}/\eta_{out}$, side branched patterns (Supplementary Movie 1) are observed. In contrast to the thick finger-like patterns reported for low $\eta_{in}/\eta_{out}$ in a pair of miscible Newtonian fluids~\cite{Lajeunesse,Irmgard}, we observe long fractal patterns with side branches under similar conditions of flow rate and viscosity ratios. As $\eta_{in}/\eta_{out}$ is increased, a dark inner circular region, devoid of fingers and comprising only the inner fluid, appears and grows larger (Fig.~\ref{sizeratio-EP}b-f). At the highest $\eta_{in}/\eta_{out}$, this inner circular region covers almost the entire pattern (Fig.~\ref{sizeratio-EP}f). Bischofberger $et$ $al$.~\cite{Irmgard} had reported the complete suppression of instabilities for a pair of Newtonian fluids at $\eta_{in}/\eta_{out}\approx\num{3.3e-1}$. We note the formation of small protrusions at the periphery of the pattern even at $\eta_{in}/\eta_{out}=\num{5.0e-1}$.

\par 
Previous numerical and experimental studies with a pair of Newtonian and non-Newtonian fluids have reported a reduction in the width of the fingers and a high dominant wavenumber at the onset of instability~\cite{Shokri_2017,Lindner_2000,Mora}. A breakdown of the interface into multiple protrusions has also been noted~\cite{Malhotra_2014}. In order to study the impact of elasticity on instabilities, we have performed experiments with cornstarch suspensions of different concentrations at a fixed $\eta_{in}/\eta_{out}$ (Supplementary Fig.~1). We observe a suppression in the growth of fingers at the periphery of the patterns and an increase in the pattern density as the concentration of the cornstarch suspensions is increased (supplementary information section S1). While the instability at the interface between a pair of Newtonian fluids depends on $\eta_{in}/\eta_{out}$~\cite{Irmgard}, our results suggest that the morphological evolution at the interface between a Newtonian and a non-Newtonian fluid is extremely sensitive to the elasticity of the outer fluid even when $\eta_{in}/\eta_{out}$ is constant.
	
\par
We follow the protocol adopted by Bischofberger $et$ $al$.~\cite{Irmgard} and quantify the patterns using different length scales (inset of Fig.~\ref{sizeratio-EP}g). $R_o$ is the radius of the smallest circle enclosing the entire pattern, $R_i$ is the radius of the largest circle enclosing only the dark region comprising the inner dyed fluid and $R_f$ is the finger length defined as $R_f \equiv R_o - R_i$. Figure~\ref{sizeratio-EP}g shows the variation of the size ratios, $R_f/R_i$, $vs.$ $\eta_{in}/\eta_{out}$ for fixed $R_{o} = 13.5$ cm. Our analysis shows that $R_f/R_i$ decreases monotonically with an increase in viscosity ratio (Fig.~\ref{sizeratio-EP}g) and resembles the observation for a pair of Newtonian fluids~\cite{Irmgard}. As $\eta_{in}/\eta_{out}$ is increased, patterns with small fingers and dark inner regions (Fig.~\ref{sizeratio-EP}b-d) that are followed by more stable interfaces with small protrusions at the periphery (Fig.~\ref{sizeratio-EP}e,f), are observed. This leads to a flatter $R_f/R_i$ curve at high $\eta_{in}/\eta_{out}$. Partly, the stable regime can be attributed to the stabilising effect of non-vanishing normal stress differences when either the displacing or the displaced fluid is viscoelastic.
	
\par
The sweep efficiency measures the significant displacement of the outer fluid by the inner fluid. Here, following the work of Shokri $et$ $al$.~\cite{Shokri_2017}, sweep efficiency is defined as the ratio of the area contacted by the inner fluid (gray area in Fig.~\ref{Viscosity-EP}a) and the total area occupied by both inner and outer fluids (sum of the black and gray areas in the same figure). Figure~\ref{sizeratio-EP}h shows the variation of sweep efficiency with time for various $\eta_{in}/\eta_{out}$. At early times, the absence of finger formation results in sweep efficiency close to one. As the fingers grow, sweep efficiency gradually decreases with time. While numerical simulations have predicted an increase in sweep efficiency with increasing elasticity of the outer fluid~\cite{Shokri_2017}, we observe an increase in sweep efficiency with increasing $\eta_{in}/\eta_{out}$ even for fixed elasticity of the fluid. 
	
\begin{figure}[!t]
\centering
\includegraphics[width=6.0in]{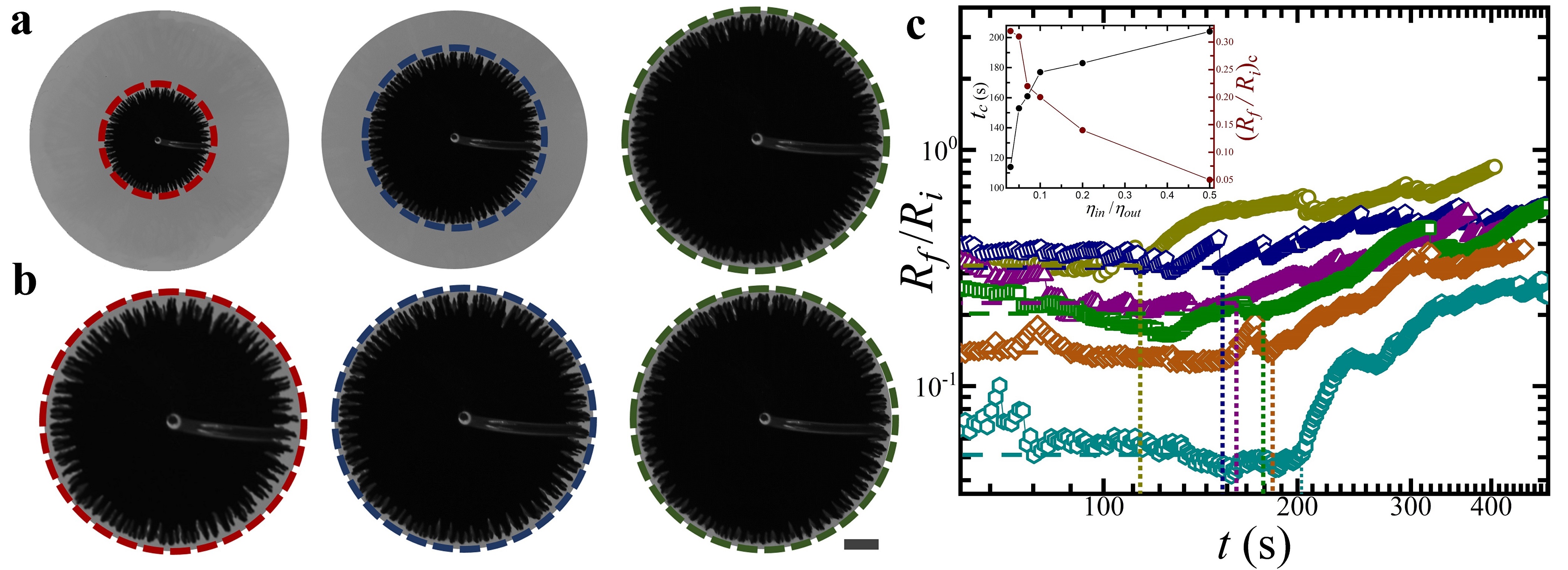}
\caption{$\vert${\bf Temporal evolution of interfacial patterns.} {\bf a,} Temporal evolution of interfacial  patterns for $\eta_{in}/\eta_{out}=\num{7.0e-2}$ (from left to right) at 61s, 81s and 101s. The scale bar is 1 cm. The coloured dashed circles enclose the inner fluid as the interfacial patterns develop with time. {\bf b,} Zoomed images of the patterns ({\bf a}) in which only the region within the coloured circle is displayed. The images are nearly indistinguishable, indicating that fingers at the periphery of the pattern grow proportionally with the overall pattern. {\bf c,} $R_f/R_i$ $vs.$ $t$ is plotted for $\eta_{in}/\eta_{out}=\num{3.0e-2}$ (\brass), $\num{5.0e-2}$ (\bluepentagon), $\num{7.0e-2}$ (\purplepentagon), $\num{1.0e-1}$ (\olivedot), $\num{2.0e-1}$ (\brown) and $\num{5.0e-1}$ (\JG) to focus on the seemingly proportionate growth of the patterns. The patterns evolve slowly ($R_f$ grows proportionally to $R_i$) at initial times ($t \leq t_{c}$). Characteristic times ($t_c$) and size ratios ($(R_f/R_i)_c$)  are shown by vertical dotted lines and horizontal dashed lines, respectively. Inset displays $t_c$ (\blbullet) and $(R_f/R_i)_c$ (\wine) $vs.$ $\eta_{in}/\eta_{out}$.}
		\label{PG-EP} 
\end{figure}

\par
We next investigate the temporal evolution of the patterns with increasing $\eta_{in}/\eta_{out}$. For $\eta_{in}/\eta_{out}$ values lying between $\num{3.0e-2}-\num{5.0e-1}$, the interface is initially stable but later develops unstable regions. Figure~\ref{PG-EP}a shows the patterns obtained for $\eta_{in}/\eta_{out}=\num{7.0e-2}$ at times $t$ = 61 s, 81 s and 101 s (Supplementary Movie 2). Circles are drawn with coloured dashed lines in Figure~\ref{PG-EP}a to enclose the inner dyed fluid completely. Figure~\ref{PG-EP}b shows magnified images of the circled regions. These patterns appear indistinguishable and indicate proportionate growth~\cite{Sadhu,Dhar}, reminiscent of the observations for Newtonian fluids~\cite{Irmgard}. They are distinct from the growing structures observed in diffusion-limited aggregation \cite{DLA, Daccord} and viscous fingering~\cite{Chen, Paterson_1981} in which growth occurs at the outer surface while inner parts, once formed, remain preserved. Figure~\ref{PG-EP}c displays $R_f/R_i$ $vs.$ time ($t$) in the regime $\eta_{in}/\eta_{out}=\num{3.0e-2}-\num{5.0e-1}$ which may be aptly termed the "proportionate growth" regime. $R_f/R_i$ remains approximately constant up to a characteristic time, $t_c$, shown by dotted vertical lines. At $t=t_c$, $R_f/R_i$ is denoted by $(R_f/R_i)_c$, shown by dashed horizontal lines in Figure~\ref{PG-EP}c and compared for different $\eta_{in}/\eta_{out}$ in the inset. For $t>t_c$,  $R_f/R_i$ starts increasing, indicating that the tip of the interface ($R_f$) grows at a faster rate than the dark inner region ($R_i$). As higher viscosity ratios are approached, the increase in $t_c$ indicates a delay in the onset of instabilities that is accompanied by a suppression of finger growth. The time-dependence of $R_f/R_i$ for the entire range of viscosity ratios is provided in the supplementary information (Supplementary Fig.~2).  
	
\begin{figure}[!t]
\centering
\includegraphics[width=4in]{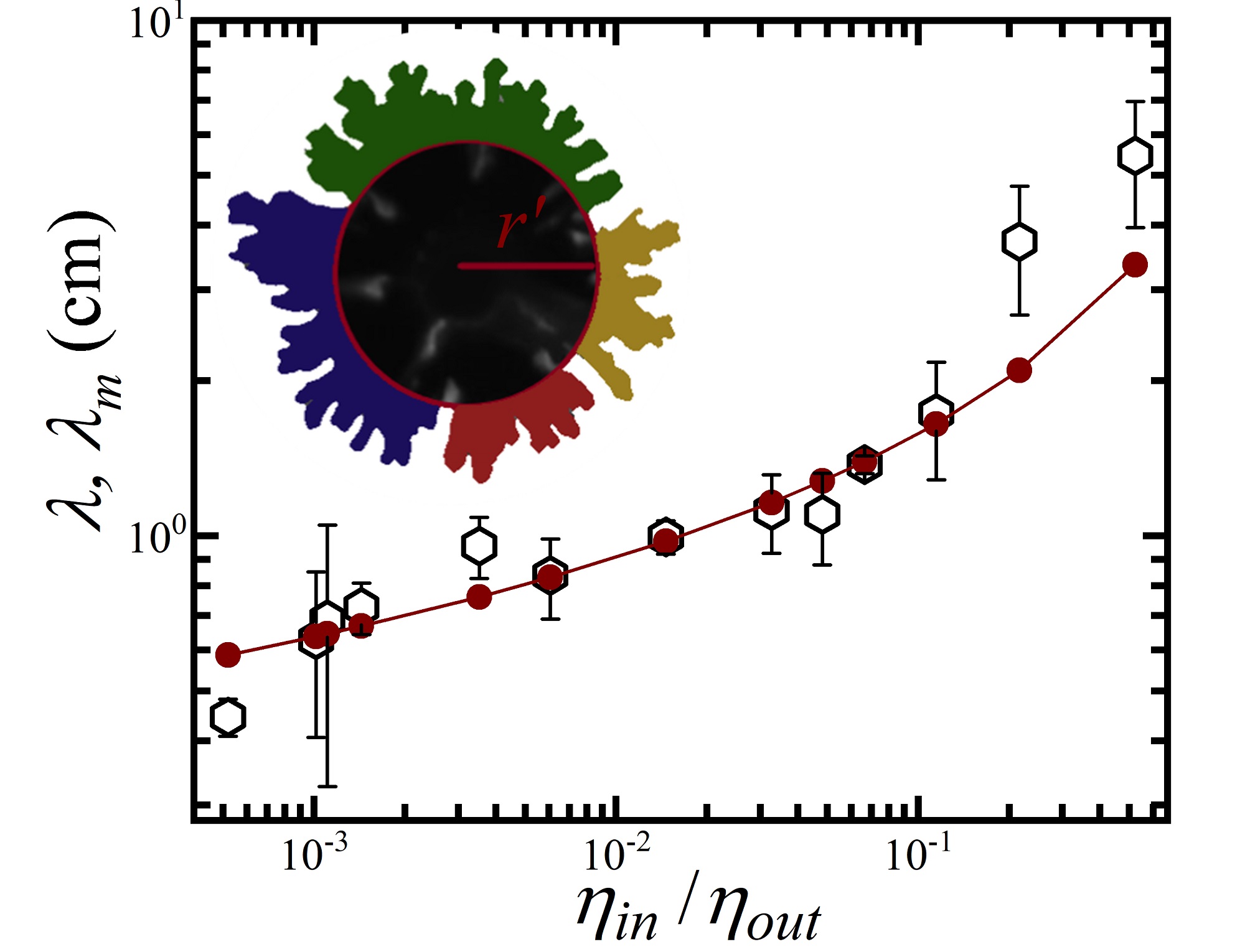}
\caption{$\vert${\bf Comparison of experimental and simulation results.} Most dominant wavelength for $\eta_{in}/\eta_{out} = \num{5.2e-4} -\num{5.0e-1}$ obtained from computational, $\lambda_{m}$  (\candyred), and experimental, $\lambda$ (\blhex), analyses for $De=0.24$, $f=2$ and $D/U =\num{1.0e-1}$ cm. Error bars represent standard deviation. Inset demonstrates the interfacial pattern at an early stage ($t$ = 5 s) in a Hele-Shaw experiment for  $\eta_{in}/\eta_{out}=\num{1.1e-3}$. A red circle of radius $r^{\prime}$ which cuts across the primary fingers (indicated by multiple colours) in the pattern is drawn. In this example, there are four primary fingers.}   
\label{lambda-EP}
\end{figure}

\par 
We compare the experimentally observed growth of the interfacial patterns with the predictions of a mathematical model~\cite{Azaiez}. The average finger spacing for different $\eta_{in}/\eta_{out}$ is estimated through pictorial measurements from experimental images at the initial stages of pattern evolution. These patterns are preferred since the model is based on linear stability analysis and ignores nonlinear effects and time-dependent viscosity changes of cornstarch suspensions that would ensue later. Initially, primary fingers on the pattern are identified qualitatively by recognising the smallest circle of radius $r^{\prime}$ that cuts across the deepest trough in the pattern. Inset of Figure~\ref{lambda-EP} displays the  experimental pattern observed at an initial stage ($t$ = 5 s) for $\eta_{in}/\eta_{out} = \num{1.1e-3}$, with the primary fingers designated by multiple colours. The average finger spacing, $\lambda$, is estimated as $\lambda = 2\pi r^{\prime}/N$, where $N$ is the number of primary fingers. This procedure is repeated for all the images obtained for different $\eta_{in}/\eta_{out}$. The values of $\lambda$, thus estimated, are compared with the dominant interfacial instability wavelength computed through linear stability analysis of the interface between a Newtonian and a shear-thinning fluid \cite{Azaiez}. We note that though the shear rate, in reality, varies with time during displacement of the outer fluid (supplementary information section S2), the shear rates of interest in this work lie in the shear-thinning regime of Figure~\ref{Viscosity-EP}b.
	
\par
An expression for the growth rate of the interfacial perturbation in terms of the perturbation wavelength (or wavenumber) and other liquid parameters is obtained from the equations governing the flow: Darcy’s law \cite{Darcy}, continuity equation and convection-diffusion equation~\cite{homsy}. Details of derivation of the growth rate are provided in the supplementary information section S3. The wavenumber (i.e., 2$\pi$/wavelength) above which the interface is stable and perturbations do not grow with time is termed the cut-off wavenumber, $k_c$, and is obtained by setting the growth rate to 0. The expression for the dimensionless form of $k_c$ is~\cite{Azaiez}:   
\begin{equation}
k_{c} = \tilde{R}{\frac{2\alpha + \zeta(1+\beta)}{(2+\beta)(1+\beta)(\alpha+\zeta)+2\beta(\alpha\beta + \zeta)}}
\label{kc}
\end{equation}
where the symbols $\tilde{R}$, $\alpha$, $\beta$ and $\zeta$  (supplementary information section S3) are shorthand notations for expressions that depend on the ratio $\eta_{out}^{o}/\eta_{in}$, Deborah number ($De$) and the exponent $n$ of the Carreau model fit in Figure~\ref{Viscosity-EP}b.
	
\par
Using equation~(\ref{kc}), the fastest-growing wavenumber, $k_{m} = k_{c}/f$ with $f = 2$ (supplementary information section S3) corresponding to the fastest growing interfacial perturbation, is computed. The most dominant wavelength of the interfacial perturbation, $\lambda_{m} = 2\pi/k_{m}$ whose dimensional values ($\lambda_m$ multiplied by $D/U$, where $D$ is the diffusion coefficient of the system and $U$ is a characteristic velocity of the fluid in the Hele-Shaw cell) are plotted in Figure~\ref{lambda-EP}. The inputs required for estimation of $k_{m}$ are $n$, $De$ and $\eta_{in}/\eta_{out}^{o}$. We set  $\eta_{out}^{o} = 3.952$ Pa.s and $n=0.32$ obtained from the fit in Figure~\ref{Viscosity-EP}b, while the other input values used are $De=0.24$ and $D/U = \num{1.0e-1}$ cm, the details for which are provided in the supplementary information section S4.
	
\begin{figure}[!t]
\centering
\includegraphics[width=5.5in]{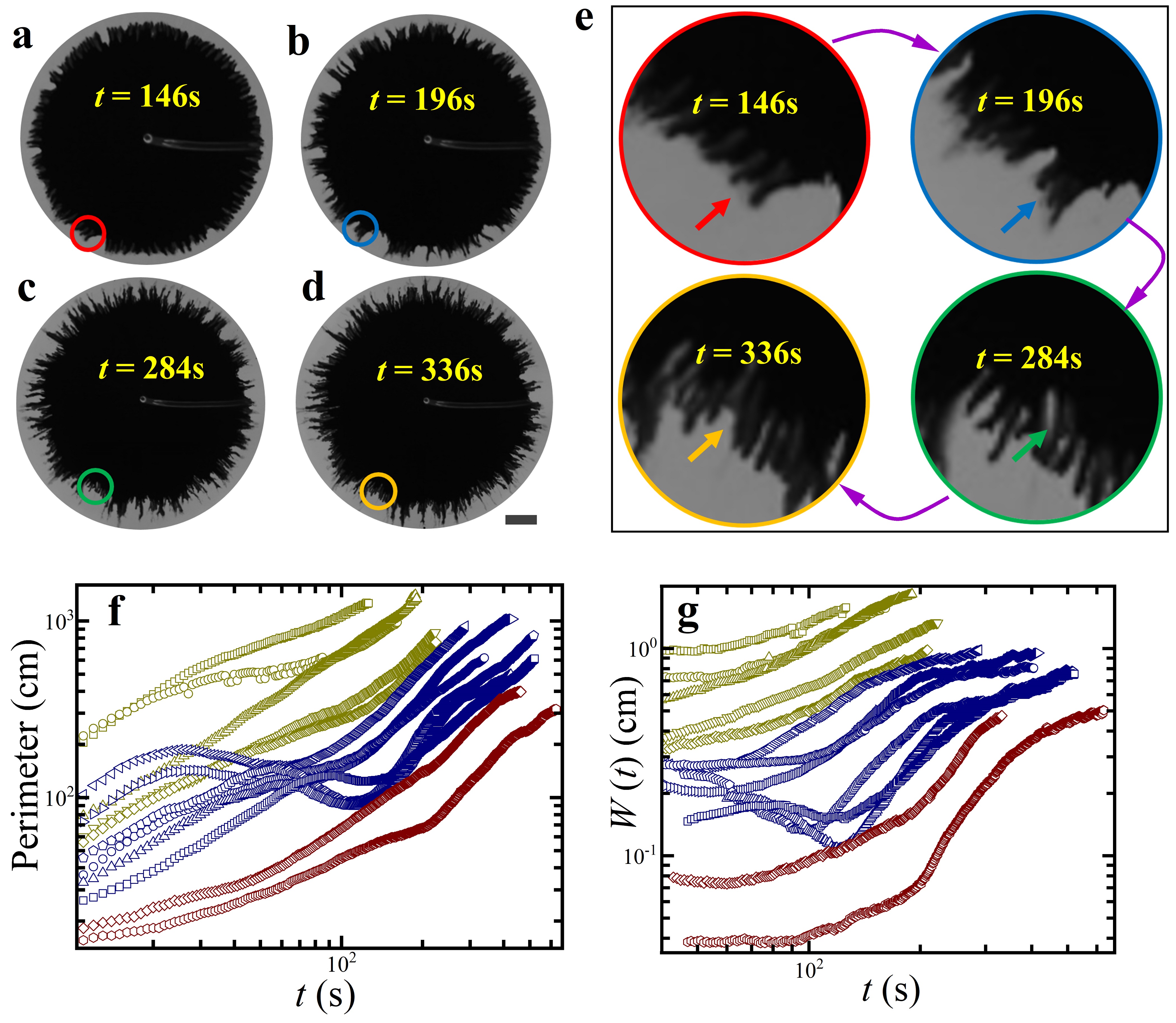}
\caption{$\vert${\bf Coalescence at high viscosity ratios.} {\bf a-d,} The temporal evolution of patterns for $\eta_{in}/\eta_{out}=\num{7.0e-2}$. Solid red, blue, green and yellow circles are drawn in {\bf a}, {\bf b}, {\bf c} and {\bf d} respectively to highlight the areas where coalescence events are observed. The scale bar is 1 cm. {\bf e,} Zoomed images of the areas within the coloured circles drawn in {\bf a-d}. Coloured arrows are drawn, pointing toward the coalescence events. The solid red circle encloses a portion of the interface just before the first coalescence event. The portions enclosed by the solid blue and green circles show the first and the second coalescence events, respectively. This is followed by the growth of new fingers enclosed by the solid yellow circle. {\bf f,} Perimeters of the patterns are plotted as a function of time for three different viscosity ratio regimes: $\num{5.2e-4}-\num{3.0e-3}$ (\brsquare,\brcircle,\brtria,\brdown,\brdiamond), $\num{6.0e-3}-\num{1.0e-1}$ (\bltriangle,\blrhd,\blcirc,\blpentagon,\bltri,\blsq) and $\num{2.0e-1}-\num{5.0e-1}$ (\rddiam,\rdhexagon). {\bf g,} Growth of overall width $W (t)$ of the interface for the same viscosity ratio regimes as indicated in {\bf f}.}
\label{coalescence-EP}
\end{figure}
		
A comparison of $\lambda_{m}$, obtained from numerical analysis, and $\lambda$, estimated from experimental data, for different $\eta_{in}/\eta_{out}$ shows good correlation (Fig.~\ref{lambda-EP}). This suggests that the most dominant wavelength of the interfacial perturbation is a reliable estimate of the average spacing between experimentally obtained finger patterns that subsequently develop. The successful application of the Carreau model-based linear stability analysis confirms the significant contribution of the shear-thinning nature of the outer fluid in determining the morphology and length scales of the patterns. The shear-thinning behaviour of cornstarch suspensions provides an additional source for the production of vorticity disturbance leading to a destabilising effect on the flow. The interfacial instability is then determined by a competition between this destabilising source and momentum dissipation that increases with increasing viscosity ratios~\cite{Azaiez}.
		
\par 
Among striking dynamics observed during the temporal evolution of the patterns is an intriguing coalescence mechanism. This is observed for $\num{6.0e-3}\leq\eta_{in}/\eta_{out}\leq\num{1.0e-1}$ that lies between the regimes of side branched patterns ($\num{5.2e-4}\leq\eta_{in}/\eta_{out}\leq\num{3.0e-3}$) and more stable interfaces ($\num{2.0e-1}\leq\eta_{in}/\eta_{out}\leq\num{5.0e-1}$). Figure~\ref{coalescence-EP}a-d shows the evolution of the patterns for $\eta_{in}/\eta_{out}=\num{7.0e-2}$ at $t$ = 146 s, 196 s, 284 s and 336 s (Supplementary Movie 3). Figure~\ref{coalescence-EP}e shows the zoomed images of areas within the coloured solid circles shown in Figure~\ref{coalescence-EP}a-d. In certain interfacial regions, the tip of a finger bends towards the adjacent finger (indicated by red arrow in Fig.~\ref{coalescence-EP}e at $t$ = 146 s). The two fingers eventually merge to form a wider finger (blue arrow in Fig.~\ref{coalescence-EP}e at $t$ = 196 s). This is followed by a second coalescence event in which a portion of the outer fluid forms a lobe that diffuses gradually in the inner fluid (green arrow in Fig.~\ref{coalescence-EP}e at $t$ = 284 s) with the subsequent formation of new fingers (yellow arrow in Fig.~\ref{coalescence-EP}e at $t$ = 336 s). To quantify the patterns in Figure~\ref{coalescence-EP}a-d, their perimeters were calculated by labelling the interface between the two fluids and counting the boundary pixels (Fig.~\ref{coalescence-EP}f). The merging of fingers in coalescence events leads to a decrease in the perimeter for intermediate $\eta_{in}/\eta_{out}$ at intermediate times. To characterise the roughness of the observed interface, the fluctuation in its radial extent is computed in terms of its overall width $W (t) = \sqrt{<[r(s,t) - <r>]^2>}$, where $r(s,t)$ is the local radius of the interface and $<...>$ denotes the average along the interface over the lateral coordinate $s$~\cite{roughness}. Coalescence events give rise to reduction in $W(t)$ for $\num{6.0e-3}\leq\eta_{in}/\eta_{out}\leq\num{1.0e-1}$ as shown in Figure~\ref{coalescence-EP}g. This is followed by an increase due to growth of new fingers at later times. Double coalescence events, also observed for other viscosity ratios $\eta_{in}/\eta_{out}=\num{3.0e-2}$ and $\eta_{in}/\eta_{out}=\num{1.0e-1}$ (Supplementary Fig.~5), were predicted by nonlinear simulations of the displacement of a viscoelastic fluid by a Newtonian fluid~\cite{Shokri_2017}, though never demonstrated before experimentally.
	
\section{Conclusions} 
We have studied interfacial instabilities between a viscoelastic outer fluid (a density matched aqueous cornstarch suspension) and a miscible inner Newtonian fluid (a glycerol-water mixture) over a wide range of $\eta_{in}/\eta_{out}$ in radial Hele-Shaw experiments. In contrast to the finger-like patterns observed in previous studies with a pair of Newtonian fluids, large scale side branched patterns are observed at low $\eta_{in}/\eta_{out}$. The elasticity of the cornstarch suspension results in the breakage of the interface into multiple protrusions. Despite the low interfacial tension, the patterns become progressively less branched with shorter fingers and display features of proportionate growth as $\eta_{in}/\eta_{out}$ increases. The experimental data is explained by a mathematical model that analyses the stability of the interface between a Newtonian and a shear-thinning viscoelastic fluid over three decades of $\eta_{in}/\eta_{out}$. Finally, an intriguing new double coalescence mechanism, predicted in nonlinear simulations of the interface between a viscoelastic and a Newtonian fluid, is observed experimentally for the first time. The elasticity of the viscoelastic outer fluid stabilises the flow, modifies the roughness of the interface and has a significant contribution in determining the morphology and time-evolution of the observed patterns. Detailed studies of flow displacements by tuning material properties of viscoelastic suspensions are important not only for applications involving large scale material processing but also from a fundamental physics point of view.

\section{Methods}
\subsection{Sample Preparation}
Inset of Figure~\ref{Viscosity-EP}b shows a cryogenic scanning electron microscopy (cryo-SEM) image of cornstarch particles acquired using a field effect scanning electron microscope (FESEM, Carl Zeiss, Germany) with an electron beam strength of 5 kV. Cornstarch powder, cesium chloride (ReagentPlus\textsuperscript{\textregistered}, 99.9\%), and rhodamine B ($\geq$95\% (HPLC)) are procured from Sigma-Aldrich and used as received. Glycerol (AR) is procured from S D Fine Chemicals Limited, Mumbai, and used as received to prepare glycerol-water mixtures. In our experiments, aqueous cornstarch suspensions (outer fluid) of viscosity, $\eta_{out}$, have been used as the non-Newtonian fluid, while glycerol-water mixtures (inner fluid) of viscosity, $\eta_{in}$, are used as the Newtonian fluid. An appropriate amount of cesium chloride (CsCl) is added to double distilled Millipore water (Millipore Corp., resistivity 18.2 M$\Omega$.cm) to prepare a 55 wt.\% CsCl solution to ensure density matching of cornstarch particles having a density of 1.59 $\mathrm{g/cm^3}$~\cite{Merkt_2004,Brown}. Cornstarch suspensions are prepared by gradually adding cornstarch powder to the cesium chloride solution. The samples are stirred thoroughly using a magnetic stirrer (1 MLH, Remi Equipments Ltd., Mumbai) for 10 minutes, followed by ultra-sonication (USC 400, ANM Industries Pvt. Ltd.) for 15 minutes to break up any large aggregates. The sample is left undisturbed for 24 hours to ensure uniform hydration of the cornstarch grains~\cite{Crawford}. The samples are re-stirred and then sonicated before each experiment to ensure sample homogeneity. The glycerol-water mixture, dyed by gradually adding rhodamine B, is stirred vigorously to ensure homogeneity. All experiments are performed at room temperature (25$^{\circ}$C).
\subsection{Hele-Shaw Cell:} 
The interfacial patterns are studied in a radial Hele-Shaw cell having two circular glass plates of radius $L$ = 30 cm with a gap $b$ = 170 $\mu$m. Teflon spacers are used to ensure a constant gap between the plates. Figure~\ref{Viscosity-EP}a shows a schematic diagram of the Hele-Shaw cell used in the experiments. The cornstarch suspension is first introduced through a 3 mm hole drilled in the top plate with a syringe pump (NE-8000, New Era Pump Systems, USA) at a constant volumetric flow rate of 2 ml/min. The glycerol-water mixture is next pumped into the cell through the same hole at a flow rate of 1 ml/min. Glycerol and water are miscible in all proportions, which allows us to control the viscosity of the mixture over a wide range. Patterns are recorded with the help of a DSLR camera set up (D5200, Nikon, Japan) below the cell. All the images are recorded at a spatial resolution of 6000 $\times$ 4000 pixels with a time interval of 1 s between successive images. The images are converted to gray-scale and analysed using the MATLAB$@$2018 toolbox. 

\subsection{Rheology}
Flow curve (viscosity) measurements are performed using an Anton Paar MCR $702$ rheometer. Steady state flow experiments of cornstarch suspensions are performed using a parallel plate geometry (PP-50) in a separate motor transducer mode. The plates each have a diameter of 50 mm and the gap between the plates is fixed at 300 $\mu$m. The sample preparation protocol for the rheological measurements is identical to that for the Hele-Shaw experiments. The sample temperature is maintained at 25$^{\circ}$C by a Peltier temperature control device (C/P-PTD 180/MD ) equipped with a water circulation unit. The sample of volume 0.5 ml is loaded between the plates. Silicone oil of viscosity 5 cSt is used as solvent trap oil to prevent evaporation. The viscosities of cornstarch suspensions of different concentrations are measured by varying the shear rate, $\dot{\gamma}$, from $0.001$ $\mathrm{s^{-1}}$ to $1000$ $\mathrm{s^{-1}}$. Shear rate $\dot\gamma$ on the cornstarch suspension imposed at the time of injection is estimated from the ratio of the flow rate of the glycerol-water mixture (1 ml/min) and characteristic volume (0.2 ml). The viscosity corresponding to this shear rate ($\dot{\gamma} = 0.08$ $\mathrm{s^{-1}}$ ) is $\eta_{out}$. Oscillatory tests are performed at a fixed angular frequency 1 rad/s over a range of strain amplitudes between $0.001$ $\%$ and $100$ $\%$ for cornstarch suspensions of different concentrations in an Anton Paar MCR $501$ rheometer. These measurements show that the elastic modulus ($G^\prime$) of dense cornstarch suspensions is higher than the viscous modulus ($G^{\prime\prime}$) at low strains, thereby signifying viscoelastic solid-like behaviour under these conditions (Supplementary Fig.~6). As strain is increased, $G^{\prime\prime}$ exceeds $G^\prime$, indicating the shear-induced transformation of the sample into a viscoelastic liquid. Furthermore, we see that the elastic modulus increases with increase in concentration of the cornstarch suspensions.

\par
A double gap geometry (DG-26.7) with an effective length of 40 mm and a gap of 1.886 mm is used for measuring the viscosity of glycerol-water mixtures in an Anton Paar MCR $501$ rheometer. The sample of volume 3.8 ml is loaded in the cell. The viscosities of the glycerol-water mixtures of different concentrations are measured while varying shear rates over three decades (Supplementary Fig.~7). The viscosity of the glycerol-water mixture, $\eta_{in}$, does not change with shear rates, which is indicative of Newtonian behaviour.

\section{Data availability}
Source data are available for this paper from the corresponding author upon reasonable request.

\section{Acknowledgements}
We thank Sayantan Majumdar, Abhishek Dhar, Joseph Samuel and Supurna Sinha for their valuable comments and feedback. We thank K. M. Yatheendran for his help with cryo-SEM imaging. We acknowledge P.G. Senapathy Computing Center, IIT Madras, for access to their computing resources.
Funding: We thank RRI for funding our research and DST SERB grant EMR/2016/006757 for partial support.
\section{Author contribution}
Palak and R.B. designed the research; Palak performed the experiments and analysed the data; R.S. \& S.K.K. performed the simulations; Palak \& R.B. drafted the manuscript, and all authors were involved in proofreading and editing.
\section{Competing Interests}
The authors declare no competing interests.
\section{Additional information}
Correspondence and requests for materials should be addressed to R.B. Supplementary Information is available for this paper. Supplementary Movie 1 shows the growth of an interfacial pattern recorded for $\eta_{in}/\eta_{out}=\num{5.2e-4}$. Supplementary Movie 2 shows the proportionate growth of an interfacial pattern recorded for  $\eta_{in}/\eta_{out} = \num{7.0e-2}$.  Supplementary Movie 3 shows the coalescence event recorded for $\eta_{in}/\eta_{out} = \num{7.0e-2}$.

\end{document}